\documentclass[a4paper,fleqn]{cas-sc}

\usepackage[numbers,sort&compress]{natbib}
\usepackage{bbding}
\usepackage{subcaption}
\usepackage{cite}
\usepackage{amsmath,amssymb,amsfonts}
\usepackage{algorithmic}
\usepackage{graphicx}
\usepackage{textcomp}
\usepackage{xcolor}
\usepackage{cuted}
\usepackage{float}
\usepackage{nomencl}
\usepackage{framed}
\usepackage{multicol}
\usepackage{threeparttable}

\def\tsc#1{\csdef{#1}{\textsc{\lowercase{#1}}\xspace}}
\tsc{WGM}
\tsc{QE}

\begin{document}
\ExplSyntaxOn
\cs_gset:Npn \__first_footerline: {}
\ExplSyntaxOff

\let\WriteBookmarks\relax
\def\floatpagepagefraction{1}
\def\textpagefraction{.001}

\shorttitle{}    

\shortauthors{G. Fu et~al.}  

\title [mode = title]{Fast Forward and Inverse Thermal Modeling for Parameter Estimation of Multi-Layer composites - Part I: Forward Modeling}  



%

\author{Gan Fu}[orcid=0009-0007-6899-1209]



\ead{g.fu@tue.nl}

\author{Calina Ciuhu}
\author{Mitrofan Curti}
\author{Elena A. Lomonova}
\affiliation{organization={Department of Electrical Engineering, Eindhoven University of Technology, 5612 AZ Eindhoven, The Netherlands}}






\cortext[1]{Corresponding author}



\begin{abstract}
This study presents fast and accurate analytical methods for transient thermal modeling in multi-layer composites with an arbitrary number of layers. The proposed approach accounts for internal heat generation and non-homogeneities in the heat diffusion equation. The separation of variables (SOV) method is employed to decouple spatial and temporal components, enabling the determination of eigenvalues. The orthogonal expansion (OE) technique is then applied to compute Fourier coefficients using `natural' orthogonality. An analytical solution for composites with constant heat sources is developed by combining the SOV method and OE technique. Additionally, a Green’s function (GF) based approach is formulated to handle transient heat sources and other non-homogeneous conditions, including temperature-dependent thermal conductivity. The results demonstrate that the proposed method offers significantly faster computations compared to finite element (FE) methods, while maintaining high accuracy. This forward modeling approach serves as an efficient basis for inverse modeling, aimed at estimating unknown material properties and geometric deformations, which are explored in Part II of this study.
\end{abstract}



\begin{keywords}
 Multi-layer composite\sep Forward modeling \sep Internal heat generation \sep Separation of variables method\sep Orthogonal expansion technique\sep Green's function method
\end{keywords}

\maketitle








\section{Introduction}

Heat conduction problem in multi-layer composites arises in a wide range of modern engineering applications, including electronic components, electrical machines, superconducting coils, lithography processes, aerospace systems, and biomedical devices~\citep{bergman2017fundamentals}. Many of these systems experience dynamic internal heat generation and non-homogeneous boundary conditions, requiring precise evaluation of temperature and heat flux distributions across multi-layer structures. Additionally, manufacturing tolerances and unforeseen experimental impediments introduce uncertainties in the thermophysical properties of materials, complicating the thermal performance assessment. Addressing these challenges requires two key objectives: accurate thermal modeling and reliable estimation of material parameters.

Heat conduction problems are generally divided into forward and inverse problems~\citep{woodburyInverseHeatConduction2023a}. In a forward problem, the temperature distribution can be calculated by solving the heat diffusion equation, given known material properties and heat flux as a function of time. This is considered a well-posed problem. In contrast, inverse problems are generally classified as ill-posed and are widely formulated as a parameter estimation approach.~\citep{hadamard1923lectures,alifanovInverseHeatTransfer1994,colacoInverseOptimizationProblems2006}. This research is presented in two parts. The first part, detailed in this paper, focuses on developing fast and accurate forward models for temperature distribution in multi-layer composites under different working conditions. These forward models are designed to be efficiently incorporated into inverse modeling frameworks, which require extensive iterations.

Forward heat conduction problems (FHCP) can be solved using either numerical methods or analytical approaches. Numerical methods, such as Galerkin-based methods, have been applied to transient conditions in multi-layer structures~\citep{demonteTransientHeatConduction2000,haji-sheikhGreensFunctionPartitioning1990,haji-sheikhIntegralSolutionDiffusion1987,curtiGeneralFormulationMagnetostatic2018a}. While numerical methods are effective for complex geometries, nonlinear materials, and intricate boundary conditions, and offer good accuracy, they provide approximate solutions and require greater computational effort than analytical methods. Analytical methods, although more difficult to derive, especially for complex geometries, provide exact solutions with shorter computation time. Several analytical methods have been developed for specific thermal problems. For instance, the separation of variables method and 'natural' analytical approach, namely, the orthogonal expansion technique have been used to model transient temperature distributions in multi-layer composites without internal heat generation~\citep{demonteAnalyticApproachUnsteady2002,demonte2003unsteady}. The Laplace transform has also been applied to similar problems, but its application is limited to three-layer slabs with Dirichlet boundary conditions~\citep{lu2005transient}. Internal heat generation introduces non-homogeneities in the heat diffusion equation, making it challenging to solve using SOV and OE alone. 
GF-based methods are particularly effective for handling internal heat generation and non-homogeneous conditions. Solutions have been developed for two-layer, three-dimensional slabs with and without heat generation~\citep{haji-sheikhTemperatureSolutionMultidimensional2002,yanThermalCharacteristicsTwoLayered1993,siegelTransientThermalAnalysis1999}. However, the studies primarily focus on cases where heat generation occurs at the boundary. A more general GF-based approach, combined with the OE technique, has also been proposed~\citep{davidw.hahnHeatConduction2012,cole2010heatgreensfunction}, but it exhibits poor convergence near the boundaries, and more complex boundary conditions are not discussed.
Therefore, no universal method exists for transient analysis of multi-layer composites, as no single technique fully addresses all heat conduction scenarios. However, effective analytical solutions can be achieved by strategically combining existing methods. 

this research presents analytical forward modeling approaches, tested on a five-layer composite structure representing a segment of a water-cooled permanent magnet linear synchronous motor (PMLSM). The goal is to establish forward thermal models that serve as the foundation for solving inverse heat conduction problems (IHCP) aimed at parameter estimation, which is discussed in Part II of this research. 
The PMLSM is selected as a representative example due to its relevance to various heat conduction scenarios, including transient heat conduction with constant and time-varying internal heat generation, and temperature-dependent material properties.
A modified combination of the SOV method and OE technique is implemented to derive solutions for transient temperature distribution in the composite, accounting for constant distributed heat generation. Detailed derivations of `natural' orthogonality, based on the regular Sturm-Liouville theorem~\citep{braunDifferentialEquationsTheir1993}, are presented. For cases involving time-dependent heat generation, a GF-based method is applied, using GFs derived from the combined SOV-OE approach. Additionally, boundary convergence issues are addressed through a superposition method~\citep{davidw.hahnHeatConduction2012}.
To validate the proposed approach, FE models are developed for comparison. The results show that the proposed analytical models offer high accuracy and computational speed, making them well-suited for integration into inverse modeling frameworks for parameter estimation.

\section{Physical model}
\nomenclature{$k_i$}{Thermal conductivity of the $i$-th layer[$\mathrm{W/(m \cdot K)}$]}
\nomenclature{$\rho_i$}{Mass density of the $i$-th layer [$\mathrm{kg/m^3}$]}
\nomenclature{$T_i(x,t)$}{Temperature of the $i$-th layer [$\mathrm{K}$]}
\nomenclature{$cp_i$}{Specific heat capacity of the $i$-th layer [$\mathrm{J/(kg \cdot K)}$]}
\nomenclature{$t$}{Time [$\mathrm{s}$]}
\nomenclature{$x, y, z$}{Spatial coordinate [$\mathrm{mm}$]}
\nomenclature{$\dot{q}_g$}{Volumetric heat generation rate [$\mathrm{W/m^3}$]}
\nomenclature{$n$}{Number of layers}
\nomenclature{$l_i$}{Thickness of the $i$-th layer [$\mathrm{mm}$]}
\nomenclature{$\alpha$}{Thermal diffusivity [$\mathrm{m^2/s}$]}
\nomenclature{$h$}{Heat transfer coefficient [$\mathrm{W/(m^2 \cdot K)}$]}
\nomenclature{$T_{water}$}{Water temperature in the tank [$\mathrm{K}$]}
\nomenclature{$T_{0}$}{Initial/room temperature [$\mathrm{K}$]}
\nomenclature{$\mathcal{X}$}{Spatial component of temperature}
\nomenclature{$\omega$}{Homogenized $T_1$}
\nomenclature{$\mathcal{T}$}{Temporal component of temperature}
\nomenclature{$\phi$}{Non-homogeneous term of $T_1$}
\nomenclature{$\lambda$}{Separation coefficient, eigenvalue}
\nomenclature{$m$}{Serial number of eigenvalues}
\nomenclature{$a, b, c$}{Coefficients of integration}
\nomenclature{$D, E$}{Integration constants}
\nomenclature{$C$}{Fourier coefficients}
\nomenclature{$F$}{Initial conditions}
\nomenclature{$N$}{Norm}
\nomenclature{$R$}{Resistance [$\mathrm{\Omega}$]}
\nomenclature{$I$}{Current [$\mathrm{A}$]}
\nomenclature{$G$}{Green's function}
\nomenclature{$f$}{Boundary conditions}
\nomenclature{$\tau$}{Time of heat source}
\nomenclature{$\theta$}{Temperature contribution\\
-$\theta_{IC}$ from initial condition\\
-$\theta_{HG}$ from heat generation\\
-$\theta_{BCs}$ from boundary conditions\\}
\nomenclature{$\text{iter}$}{Order of iteration}
\nomenclature{$\varepsilon$}{Tolerance of temperature difference}

\begin{figure*}
    \centering
    \includegraphics[width=1\linewidth]{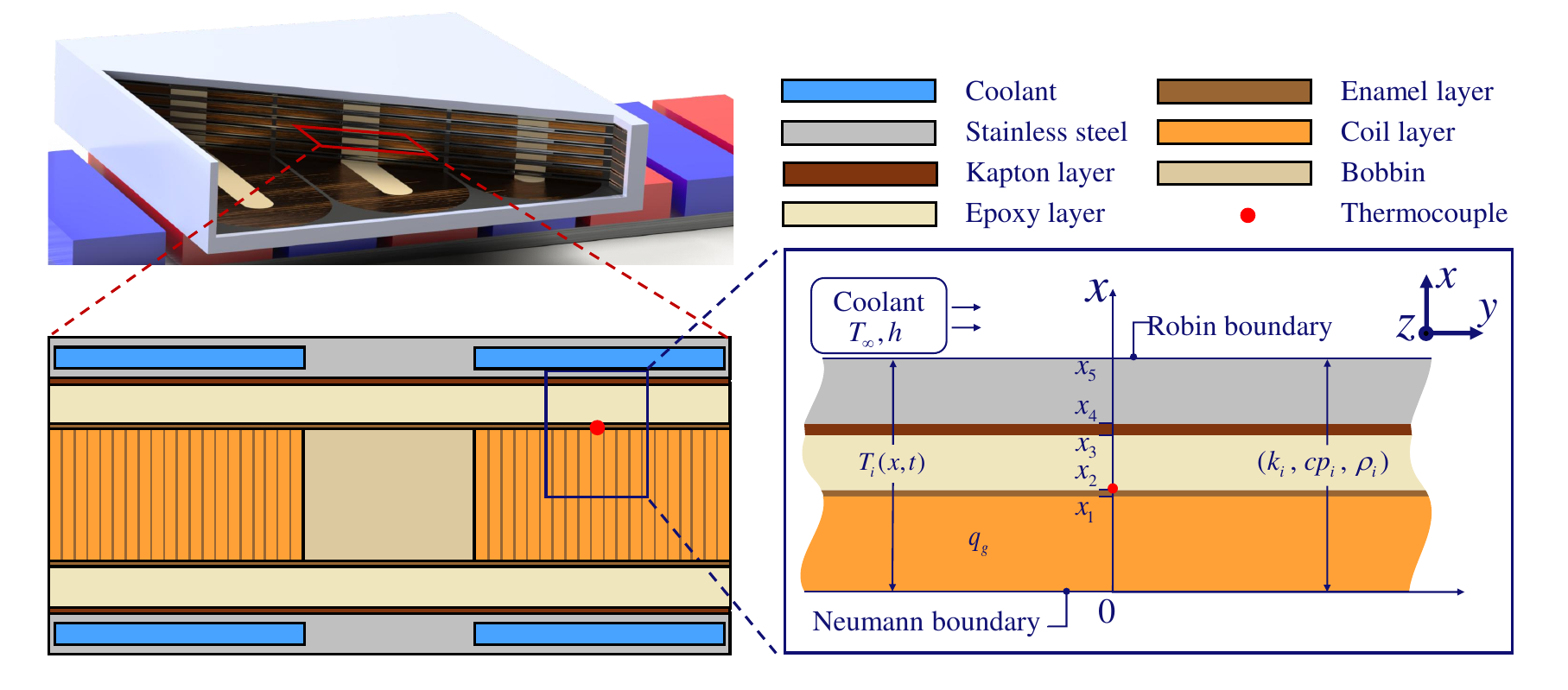}
    \caption{Five-layer composite structure from a section of the water-cooled PMLSM mover, with heat generation occurring in the coil layer. A thermocouple positioned at the enamel-epoxy interface.}
    \label{fig:LayerStructure}
\end{figure*}

A three-phase, six-layer PMLSM is used as the example in this study. In this motor, the coils are constructed from enameled foil wires and cast in epoxy to enhance thermal performance. A Kapton layer is bonded on top of the epoxy to provide additional insulation strength. Each coil is further enclosed between two cooling plates. The symmetrical and repetitive nature of this structure allows us a representative part of the motor to be selected for thermal analysis. As depicted in Fig.~\ref{fig:LayerStructure}, the analyzed section comprises five distinct material layers, representing a typical multi-layer heat conduction problem with internal heat generation.

\subsection{Heat diffusion equation}
Ohmic heating occurs in the coil layer and is conducted through all layers before being dissipated by the cooling plate. The coil is wound by turns of enameled flat wires, as shown in Fig.~\ref{fig:LayerStructure}, exhibiting high anisotropy of the thermal conductivity across different directions.
The thermal conductivity of the coil layer in the $y$-direction can be estimated using the harmonic mean of the conductivities of copper and enamel, while a weighted mean is used to calculate the effective thermal conductivity in the $x$- and $z$-directions~\citep{fuAnisotropic3DThermal2024}. This yields thermal conductivities of approximately 336 $\mathrm{W/(m \cdot K)}$ in the $x$- and $z$-directions, and 2.5 $\mathrm{W/(m \cdot K)}$ in the $y$-direction.
Given the high conductivity and dominant heat flow in the $x$-direction, the model can be simplified to a one-dimensional (1D) analysis along this axis, where the most significant temperature gradient is observed.

For a $n$-layer composite with distributed heat generation in the first layer, the 1D heat diffusion equations are formulated as:
\begin{align}
 \rho_1 cp_1 \frac{\partial T_1(x,t)}{\partial t}&=  k_1\frac{\partial^2 T_1(x,t)}{\partial x^2}+\dot{q}_g, \ \ x\in[0,x_1], 
\label{eq.multilayer_heat equation copper} \\
\rho_i cp_i \frac{\partial T_i(x,t)}{\partial t}&=  k_i\frac{\partial^2 T_i(x,t)}{\partial x^2}, \ \ x\in[x_{i-1},x_i] \ (i = 2,...,n).
\label{eq.multilayer_heat equation insulation}  
\end{align}

In this model, the coil layer is identified as layer 1, and n=5, representing the number of layers. The term $\dot{q}_g$ represents the internal heat generation rate in the coil layer, which is determined by the applied electrical current. The position of each layer, $x_i$, is defined as $x_i = \sum_1^i l_i$, where $l_i$ is the thickness of the $i$-th layer. The thermal diffusivity, $\alpha_i = k_i/(cp_i \cdot \rho_i)$, is defined for each layer, with $\rho_i$ and $cp_i$ representing the corresponding density and specific heat capacity, as summarized in Table~\ref{tab:MateirialProperties}.
\begin{table}[H]
\begin{threeparttable}
    \centering
    \caption{Thermophysical properties and thickness of the layers}
    \begin{tabular}{l l c c c c}
        \hline
        \hline
        $i$ & Layer & $k_i$ & $cp_i$ & $\rho_i$ & $l_i$ \\
        \ &  &$\mathrm{[\frac{W}{m \cdot K}]}$ & $\mathrm{[\frac{J}{kg \cdot K}]}$ & $\mathrm{[\frac{kg}{m^3}]}$ & [mm] \\
        \hline
         1 & Coil & 336 & 404 & 7734 & 0.78\\
         2 & Enamel & 0.4 & 1100 & 1300 & 0.02 \\
         3 & Epoxy & 1.3 & 800 & 2200 & 0.2 \\
         4 & Kapton & 0.4 & 1100 & 1300 & 0.025\\
         5 & Stainless steel & 16.3 &  500 & 8000 & 0.25\\ 
         \hline
         \hline
    \end{tabular}
    \label{tab:MateirialProperties}
    \begin{tablenotes}
    \footnotesize
    \item[*] The coil layer is composed of foil copper wires coated with enamel insulation.
    \end{tablenotes}
    \end{threeparttable}
    \end{table}
\subsection{Boundary and initial conditions}

As depicted in Fig.~\ref{fig:LayerStructure}, symmetrical boundary condition (Neumann type) and the cooling boundary condition (Robin type) are applied at the bottom and top edges, respectively. At each interface between layers, the continuous boundary conditions should be satisfied, ensuring that both temperature and heat flux remain continuous. These boundary conditions are expressed as: 
\begin{equation}
    \begin{split}
        \frac{\partial T_1}{\partial x}(x,t)|_{x=0}&=0,\\
        T_i(x,t)|_{x=x_i}&=T_{i+1}(x,t)|_{x=x_i}, \   (i=1,...,n-1),\\
        k_i\frac{\partial T_i(x,t)}{\partial x}\Big|_{x=x_i}&=k_{i+1}\frac{\partial T_{i+1}(x,t)}{\partial x}\Big|_{x=x_i}, \  i=(1,...,n-1), \\
        k_n \frac{\partial T_n(x,t)}{\partial x}\Big|_{x=x_n}&=-h(T_n(x,t)|_{x=x_n}-T_{\text{water}}),
    \end{split}
    \label{eq:BCs}
\end{equation}
where $T_{\text{water}}$ is the cooling water temperature, and $h$ is the heat transfer coefficient at the water-stainless steel interface, as estimated in~\citep{fuAnisotropic3DThermal2024}. 
The initial condition is assumed to be uniform throughout the composite, as in many cases, the temperature typically starts from a stable ambient value $T_i(x,0)=T_{0}$, where $T_{0}$ represents the room temperature.

\section{Constant heat source}\label{subsection_constant_qg}
To derive an exact closed-form analytical solution for the temperature distribution across all layers, the partial differential equations (PDEs) given in Eq.~\eqref{eq.multilayer_heat equation copper} and Eq.~\eqref{eq.multilayer_heat equation insulation} need to be solved. In this section, the solution is obtained from a modified combination of the SOV method and OE technique, under the assumption that $\dot{q}_g$ is constant.

\subsection{Separation of variables method}
To simplify the non-homogeneous PDE~\eqref{eq.multilayer_heat equation copper}, a transformation is applied by introducing $\omega(x,t) = T_1(x,t) - \phi(x)$. This reduces the equation to a homogeneous form~\citep{carslaw1959conduction}:
\begin{equation}
 \rho_1 cp_1\frac{\partial \omega}{\partial t}=  k_1\frac{\partial^2 \omega}{\partial x^2} \ \ x\in[0,x_1],
    \label{eq.homogenized heat equation copper}  
\end{equation}
where $\phi(x) = -\dot{q}_g x^2/2k_1$.
The temperature distribution $T_i(x,t)$ in each layer, governed by equations Eq.~\eqref{eq.multilayer_heat equation copper} and Eq.~\eqref{eq.multilayer_heat equation insulation}, can be expressed as the product of a spatial $\mathcal{X}(x)$ and a temporal $\mathcal{T}(t)$ component, following the SOV method~\citep{demonteTransientHeatConduction2000,davidw.hahnHeatConduction2012,polyaninSeparationVariablesExact2021}. This is written as:
\begin{equation}
    \begin{split}
    \omega(x,t)=\mathcal{X}_1(x)\mathcal{T}_1(t),\\
    T_i(x,t)=\mathcal{X}_i(x)\mathcal{T}_i(t).
    \end{split}   
    \label{eq:SOV method}
\end{equation}
Substituting~\eqref{eq:SOV method} into the heat equations~\eqref{eq.multilayer_heat equation insulation} and~\eqref{eq.homogenized heat equation copper}, yields a set of ordinary differential equations:
\begin{equation}
\alpha_i \frac{\mathcal{X}^{''}_i(x)}{\mathcal{X}_i(x)}  =  \frac{\mathcal{T}^{'}_i(t)}{\mathcal{T}_i(t)} = -\lambda_i^2, \  (i = 1,...,n).
\label{eq.multilayer_SV_lambda_i}
\end{equation}
The spatial components $\mathcal{X}_i(x)$ can be determined by solving the \textit{Helmholtz equations}~\citep{demonteTransientHeatConduction2000}, while solutions for the temporal components $\mathcal{T}_i(t)$ are readily obtained:
\begin{equation}
\begin{split}
    \mathcal{X}_i(x)&=a_i sin(\lambda_i x)+b_i cos(\lambda_i x), \\
    \mathcal{T}_i(t)&=c_i e^{-\lambda_i^2 \alpha_i t}.
\end{split}    
\label{eq:SOV method solutions_unmodified}
\end{equation}
To ensure the continuity conditions at layer interfaces and the uniqueness of the solution, $\lambda_i$ should satisfy: $\lambda^2_i \alpha_i = \lambda^2_{i+1} \alpha_{i+1} = \lambda^2$, i.e., $\lambda_i= \lambda/\sqrt{\alpha_i}$. While $\lambda$ is the eigenvalue of this thermal problem, also known as the separation constant~\citep{zhouTheoreticalSolutionTransient2017}. So that the solutions of the spatial and temporal components are modified as:
\begin{equation}
\begin{split}
    \mathcal{X}_i(x)&=a_i sin(\lambda x/\sqrt{\alpha_i})+b_i cos(\lambda x/\sqrt{\alpha_i}), \\
    \mathcal{T}_i(t)&=c_i e^{-\lambda^2 t}.
\end{split}    
\label{eq:SOV method solutions}
\end{equation}
The general solution for $T_i(x,t)$ can then be expressed as:
\begin{equation}
\begin{split}
    T_i(x,t)=&(a_i sin(\lambda x/\sqrt{\alpha_i})+b_i cos(\lambda x/\sqrt{\alpha_i}))c_i e^{-\lambda^2 t}\\
    &+D_i x+E_i \ (-\dot{q}_gx^2/2k_1, \  \text{when} \ i=1).
\end{split}
\label{eq: T_general}
\end{equation}
The solution~\eqref{eq: T_general} subject to the boundary conditions~\eqref{eq:BCs}, results in two systems of linear equations:
\begin{align}
    M_{DE} [D_1, E_1,..., D_n, E_n]^T&= A ,
    \label{eq,solve DsEs} \\
    M_{ab\lambda} [a_1, b_1,..., a_n, b_n]^T&= 0 .  
    \label{eq,solve AsBs}
\end{align}
The terms $D_i$ and $E_i$ are integration constants, which can be easily determined by solving the system of linear equations~\eqref{eq,solve DsEs} with $n$ elements. While $M_{ab\lambda}$ is a 2$n$-by-2$n$ matrix, which contains the unknown eigenvalue $\lambda$ that needs to be determined, written as:

\begin{equation}
    M_{ab\lambda}=
    \begin{bmatrix}
        \quad 1 \quad & 0 & \cdots & 0 \\
        & & M_{ct} & \\
        & & M_{cf} & \\
        \quad 0 \quad & \cdots & \frac{\lambda}{\sqrt{\alpha_n}} k_n \cos\left(\frac{\lambda}{\sqrt{\alpha_n}} x_n\right)+h \sin\left(\frac{\lambda}{\sqrt{\alpha_n}} x_n\right) & -\frac{\lambda}{\sqrt{\alpha_n}} k_n \sin\left(\frac{\lambda}{\sqrt{\alpha_n}} x_n\right)+h \cos\left(\frac{\lambda}{\sqrt{\alpha_n}} x_n\right)
    \end{bmatrix},
    \label{eq.multilayer_matrix}
\end{equation}
    \begin{equation}
    M_{ct} \ \ =
    \left. \begin{bmatrix}
    M_{ct}^{(1)} & \cdots & 0
    \\
    \vdots & \ddots &\vdots 
    \\
    0 \quad & \cdots & M_{ct}^{(n-1)}
    \end{bmatrix}\right|_{(n-1)\times 2n},
\hspace{50pt}
    M_{cf}=
    \left. \begin{bmatrix}
    M_{cf}^{(1)}   & ... & 0
    \\
    \vdots & \ddots &\vdots 
    \\
     0 \quad & ...&M_{cf}^{(n-1)}
    \end{bmatrix} \right|_{(n-1)\times 2n},
    \label{eq.matrix_cf}
\end{equation}
\begin{equation}
\begin{split}
    M_{ct}^{(i)} &= \left[ sin(\frac{\lambda}{\sqrt{\alpha_{i}}} x_{i}), \  cos(\frac{\lambda}{\sqrt{\alpha_{i}}} x_{i}), \  -sin(\frac{\lambda}{\sqrt{\alpha_{i+1}}} x_{i}), \  -cos(\frac{\lambda}{\sqrt{\alpha_{i+1}}} x_{i}) \right], \\
    &  \hspace{30pt} (2i-1) \hspace{30pt} (2i) \hspace{40pt} (2i+1) \hspace{40pt} (2i+2)    
\end{split}
\end{equation}
\begin{equation}
\begin{split}
    M_{cf}^{(i)} &= \left[ -\frac{\lambda}{\sqrt{\alpha_i}} k_i cos(\frac{\lambda}{\sqrt{\alpha_i}} x_i), \ \frac{\lambda}{\sqrt{\alpha_i}} k_i  sin(\frac{\lambda}{\sqrt{\alpha_i}} x_i), \ \frac{\lambda}{\sqrt{\alpha_{i+1}}} k_{i+1} cos(\frac{\lambda}{\sqrt{\alpha_{i+1}}} x_i), \  -\frac{\lambda}{\sqrt{\alpha_{i+1}}} k_{i+1} sin(\frac{\lambda}{\sqrt{\alpha_{i+1}}} x_i)\right].\\
    &  \hspace{50pt} (2i-1) \hspace{60pt} (2i) \hspace{70pt} (2i+1) \hspace{80pt} (2i+2)
\end{split}
\end{equation}
\begin{figure*}[h!]
    \centering
    \includegraphics[width=0.85\textwidth]{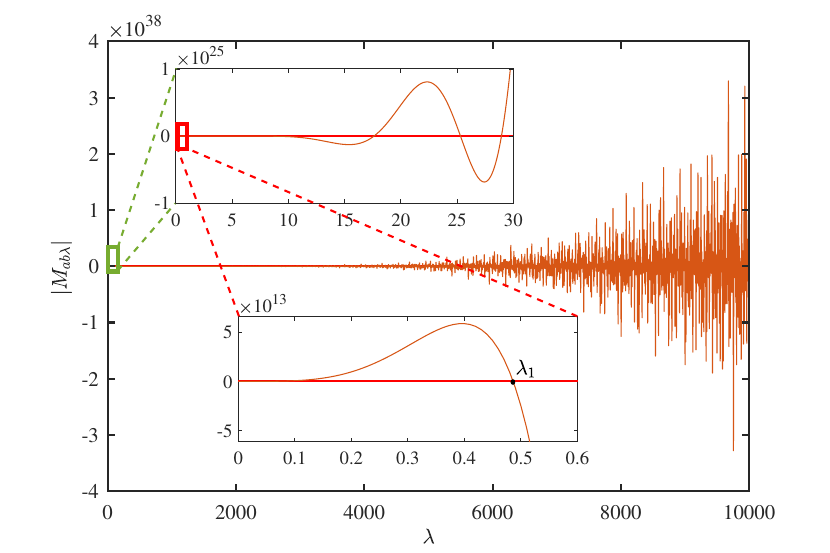}
    \caption{Graphical representation of the eigenvalues $\lambda_m$ obtained from the transcendental equation $|M_{ab\lambda}|=0$.}
    \label{fig:lambda}
\end{figure*}
\begin{figure*}[h!]
    \centering
    \includegraphics[width=1\textwidth]{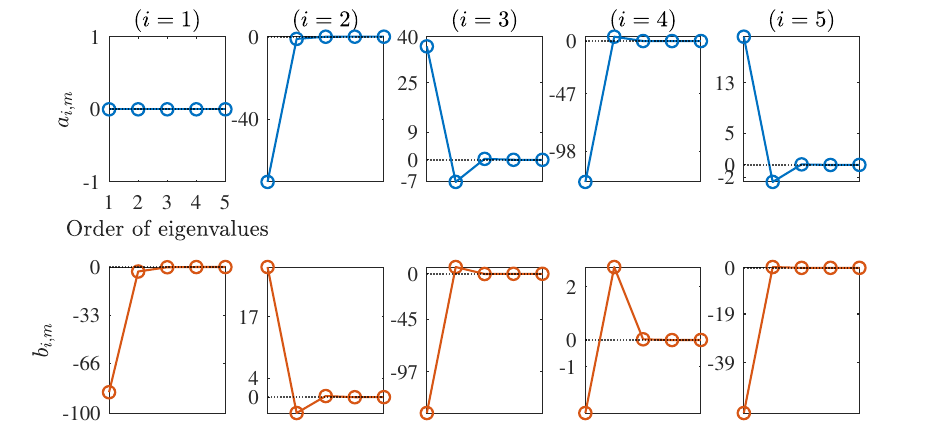}
    \caption{Convergence behavior of the coefficients $a_{i,m}$ and $b_{i,m}$ for $i = 1,\dots,5$ and $m = 1,\dots,5$.}
    \label{fig:convergence_coefficients}
\end{figure*}
Where $M_{ct}$ and $M_{cf}$ are the parts from the continuous temperature and continuous heat flux boundary conditions.
Considering the existence of a non-trivial solution, the determinant of $M_{ab\lambda}$ must be 0, i.e., $|M|=0$. This leads to a transcendental equation used to determine the eigenvalues $\lambda_m$~\citep{davidw.hahnHeatConduction2012}. Note that an infinite number of $\lambda_m$ can be found, as shown in Fig.~\ref{fig:lambda}, only the first few have a significant influence, as indicated by the convergence behavior of the coefficients in Fig.~\ref{fig:convergence_coefficients}. It is demonstrated that the first three eigenvalues are sufficient to accurately construct the temperature solution.

Since the problem forms a regular Sturm-Liouville problem, the spatial solution $\mathcal{X}_i(x)$ can be expanded into a convergent series of eigenfunctions~\citep{braunDifferentialEquationsTheir1993}, yielding the general solution of the temperature distribution $T_i(x,t)$: 
\begin{equation}
    T_i(x,t) = \sum_{m=1}^\infty C_m e^{-\lambda_m^2 t} \mathcal{X}_{i,m}(x) +D_i x+E_i \ (-\dot{q}_gx^2/2k_1, \ \text{when} \ i=1),
    \label{eq:T_sp}
\end{equation}
where:
\begin{equation}
\begin{split}
    \mathcal{X}_{i,m}(x) &\equiv \mathcal{X}_{i}(\lambda_m,x) \\
    &=a_{i,m} sin(\lambda_m x/\sqrt{\alpha_i})+b_{i,m} cos(\lambda_m x/\sqrt{\alpha_i}).     
\end{split}
\end{equation}
Here, $a_{i,m}$ and $b_{i,m}$ are coefficients from the null space of $M_{ab\lambda}$, normalized by $c_i a_{1,m}/C_m$, where $C_m$ is called the Fourier coefficient~\citep{davidw.hahnHeatConduction2012}.

\subsection{Orthogonal expansion technique}

At this stage, the only remaining unknown in the solution is the coefficient $C_m$, which can be determined by applying the initial condition. Using $T_i(x,0) = T_0$, we obtain:
\begin{equation}
\begin{split}
    F_i(x) &= T_0 - D_i x - E_i \ (+ \dot{q}_gx^2/2k_1, \ \text{when} \ i=1) \\
    &= \sum_{m=1}^\infty C_m  \mathcal{X}_{i,m}(x), 
    \label{eq:IC}    
\end{split}
\end{equation}

Since the eigenfunctions $\mathcal{X}_{i,m}(x)$ form a complete set of basis functions that satisfy the condition of `natural' orthogonality~\citep{demonteAnalyticApproachUnsteady2002}, their inner product takes the form of a Kronecker delta function:
\begin{equation}
        \sum^n_{i=1}  \int^{x_i}_{x_{i-1}}  \frac{k_i}{\alpha_i} \mathcal{X}_{i,m}(x)\mathcal{X}_{i,p}(x) dx = 
    \begin{cases}
     0&\  m  \neq p ,
     \\
     N_m &\  m = p,
 \end{cases}
 \label{eq:orthogonal merit}
\end{equation}
where the norm $N_m$ is defined as:
\begin{equation}
      N_m=  \sum^n_{i=1} \int^{x_i}_{x_{i-1}} \frac{k_i}{\alpha_i}  (\mathcal{X}_{i,m}(x))^2 dx. 
\end{equation}
By virtue of the orthogonality of the eigenfunctions, the Fourier coefficients $C_m$ can be explicitly determined from Eq.~\eqref{eq:IC} and Eq.~\eqref{eq:orthogonal merit}, as detailed in Appendix~\ref{Appendix_OET}.
\begin{equation}
 C_m=\frac{1}{N_m}\sum^n_{i=1}  \int^{x_i}_{x_{i-1}} \frac{k_i}{\alpha_i} F_i(x)\mathcal{X}_{i,m}(x) dx.
 \label{eq:fourier coefficient}
\end{equation}
Thus, the exact closed-form solution for the temperature distribution in each layer, as given by Eq.~\eqref{eq:T_sp}, can be fully assembled using the determined parameters $\lambda_m, a_{i,m}, b_{i,m}$, $C_m$ along with $D_i$ and $E_i$.

This approach is generic and can be applied to composites with an arbitrary number of layers, where constant heat sources may exist in any layer or simultaneously in multiple layers. Furthermore, different boundary and initial conditions can be incorporated with minimal adjustments. Only the matrices $M_{DE}$ and $M_{ab\lambda}$ need to be modified to reflect the specific conditions, while the overall solution procedure remains unchanged.

\subsection{Validation}

\begin{figure*}[h!]
\centering
\begin{subfigure}{0.52\textwidth}
    \includegraphics[width=\linewidth]{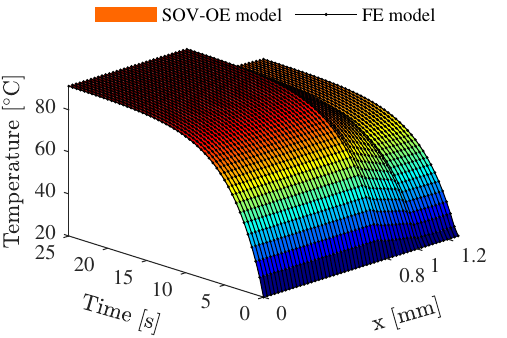}
    \caption{Temperature distribution.}
    \label{fig:T_x}
\end{subfigure}
\hfill
\begin{subfigure}{0.44\textwidth}
    \includegraphics[width=\linewidth]{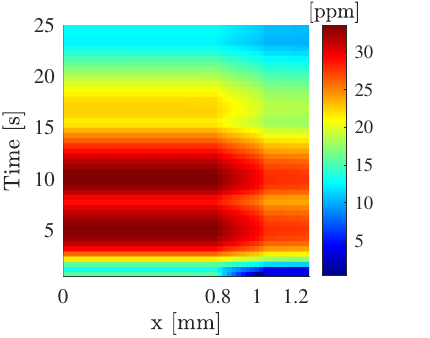}
    \caption{Error distribution.}
    \label{fig:T_t}
\end{subfigure}
\caption{Comparison of spatial and temporal temperature distributions of the multi-layer composite under constant heat generation, based on solutions from the combined SOV-OE analytical model and the 1D FE model, along with the corresponding error distribution.}
\label{fig:_T_profiles}
\end{figure*}

To validate this combined SOV-OE analytical model, a 1D FE model with identical layer configurations, boundary conditions, and initial conditions was developed using COMSOL platform. A direct current (DC) of 25 A was applied to the coil, and the corresponding internal heat generation rate $\dot{q}_g$ was calculated based on the volumetric ohmic losses, using the parameters listed in Table~\ref{tab. Working conditions}. The spatial-temporal temperature profile obtained from the FE model was compared to that derived from the SOV-OE model, as shown in Fig.~\ref{fig:_T_profiles}, along with the corresponding error distribution.
\begin{table}[h!]
\begin{threeparttable}
    \centering
    \caption{Working conditions of the PMLSM.}
    \begin{tabular}{l l l l}
        \hline
        \hline
        Symbol & Value & Unit& Description  \\
        \hline
         $R_c$ & 0.565 & $\Omega$ & Coil resistance \\
         $I_{DC}$ & 25 &A & Direct current source\\
         $\dot{q}_g$ & 7.4$\times 10^7$ & $\mathrm{W/m^3}$ & Heat generation rate\\
         $h$ & 1050 & $\mathrm{W/(m^2\cdot K)}$& Heat transfer coefficient \\
         $T_0$ & 293.15 & $\mathrm{K}$ & Initial temperature \\
         $T_{\text{water}}$ & 293.15 & $\mathrm{K}$ & Water temperature \\
         $m$ & 3 & - & Fourier coefficient number\\
         \hline
         \hline
    \end{tabular}
    \label{tab. Working conditions}
    \end{threeparttable}
\end{table}

The results demonstrate strong agreement between the SOV-OE and FE models, with a maximum error of less than 40 parts per million (ppm). 
In terms of computational efficiency, the SOV-OE model achieves a simulation time of just 0.013 s, compared to 2.5 s for the 1D FE model using the same spatial discretization and time steps on a 7980x 64-cores CPU and 384 GB of RAM. The FE method relies on a time-stepping approach, which is inherently slower and may introduce errors that accumulate over time steps. Additionally, COMSOL may contribute to time overhead, as commercial software often performs supplementary internal computations alongside the primary simulation tasks. 

\subsection{Selection of eigenvalues}
Since the number of eigenvalues used in the temperature solution directly affects both computational time and accuracy, selecting an appropriate number that balances these two aspects is important. As indicated in Fig.~\ref{fig:convergence_coefficients}, the coefficients $a_{i,m}$ and $b_{i,m}$ converge rapidly beyond the third term, already providing satisfactory accuracy. To further investigate this behavior, the relationship between the number of eigenvalues, computational cost, and solution accuracy is analyzed for the analytical SOV-OE model. Additionally, the effect of relative tolerance on the FE model accuracy is examined.

\begin{figure*}[h!]
\centering
\begin{subfigure}{0.48\textwidth}
    \includegraphics[width=\linewidth]{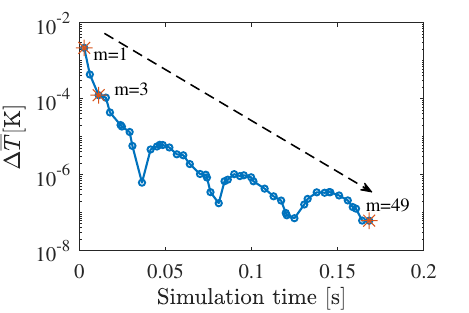}
    \caption{$\Delta \overline{T}$ versus number of eigenvalues $\lambda_m$ (SOV-OE model).}
    \label{fig:number of eigenvalues}
\end{subfigure}
\hfill
\begin{subfigure}{0.48\textwidth}
    \includegraphics[width=\linewidth]{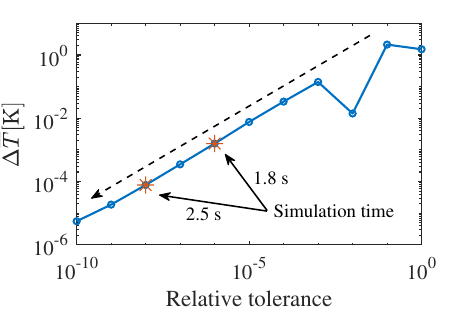}
    \caption{$\Delta \overline{T}$ versus relative tolerance (FE model).}
    \label{fig:relative tolerance}
\end{subfigure}
\caption{Convergence evaluation for determining eigenvalues: (a) effect of number of eigenvalues on SOV-OE model accuracy; (b) effect of relative tolerance on FE model accuracy. All results are referenced to $T_{m=50}(t)$.}
\label{fig:convergence of eigenvalues}
\end{figure*}

A temperature solution $T_{m=50}(t)$ obtained at the enamel-epoxy interface using 50 eigenvalues is taken as the reference, as this solution has been validated to provide high accuracy. Temperature solutions from the SOV-OE model using various numbers of eigenvalues are also compared to $T_{m=50}(t)$, as presented in Fig.~\ref{fig:number of eigenvalues}. Meanwhile, solutions from the FE model with different relative tolerance settings are compared to $T_{m=50}(t)$, as shown in Fig.\ref{fig:relative tolerance}. 
Where $\Delta \overline{T}$ represents the average temperature difference per time step between each solution and the reference $T_{m=50}(t)$. 
The results indicate that lower relative tolerances in the FE model lead to improved agreement with $T_{m=50}(t)$, at the cost of increased simulation time. A similar trend is observed for the SOV-OE model when increasing the number of eigenvalues. However, a stepwise convergence behavior appears in the analytical solution: in certain ranges, adding more eigenvalues may introduce oscillations in accuracy while still increasing simulation time. This is attributed to the nature of the transcendental eigenvalue solution process.
Notably, using $m=1$ in the SOV-OE model yields accuracy comparable to the 1D FE model with a relative tolerance of $10^{-6}$, while reducing simulation time by a factor of approximately 800. Similarly, $m=3$ achieves identical accuracy as the FE model with a tolerance of $10^{-8}$, reducing the simulation time by 200. Corresponding simulation times for both models are summarized in Table~\ref{tab:Simulation time}.
\begin{table}[h!]
    \begin{threeparttable}
    \centering
    \caption{Simulation time.}
    \begin{tabular}{L C c c c c}
        \hline
        \hline
          & SOV-OE & 1D FE & SOV-OE & 1D FE & \\
          & $m=1$ & $\mathrm{Re} = 10^{-6}$& $m=3$ & $\mathrm{Re} = 10^{-8}$ \\
         \hline
         Simulation time &0.002 s  & 1.8 s & 0.013 s  & 2.5 s \\ 
         \hline
         \hline
    \end{tabular}
    \label{tab:Simulation time}
    \begin{tablenotes}
    \footnotesize
    \item[*] The SOV-OE model has identical spacial nodes, and time steps with the 1D FE model. 
    \end{tablenotes}
    \end{threeparttable}
\end{table}

Based on this analysis, an appropriate number of eigenvalues can be selected depending on the required accuracy and computational resources. In this study, $m=3$ is adopted throughout the paper.

\section{Transient heat source}

Multi-layer composite materials are widely used due to their superior thermal and mechanical performance, making them well-suited for demanding operating environments~\citep{zhouTheoreticalSolutionTransient2017}. In many applications including electrical machines, heat generation is not constant but varies over time. Common current waveforms such as sinusoidal signals and pulse-width modulation (PWM) result in dynamic heat generation. Additionally, specific operational requirements may introduce rapidly changing current profiles, leading to time-dependent internal ohmic losses, which propagate through the entire composite structure.

Accurately modeling time-varying heat sources poses challenges when using the previously discussed SOV method and OE technique. These approaches require recalculating all coefficients at each time step as $\dot{q}_g$ varies over time, leading to potential discontinuities in the solution and increased computational cost, despite maintaining good accuracy.
Furthermore, in electrical machines, the electrical resistivity of the copper in the coil winding is temperature-dependent, influencing the ohmic losses and, consequently, the temperature distribution. Ignoring this effect can lead to significant prediction errors, particularly under dynamic conditions.

To address these challenges, a GF-based approach is adopted, building on the solution from the combined SOV-OE model. This approach provides a continuous solution without the need to update coefficients at every time step, improving the computational efficiency.

\subsection{Green's function method}
The GF method is a powerful tool to solve the PDEs with transient sources or boundary conditions. In the 1D multi-layer structure, the temperature solution can be represented as the superposition of the contributions from the initial condition (IC), heat generation (HG), and boundary conditions (BCs), formulated as~\citep{davidw.hahnHeatConduction2012,ozisik1993heat,cole2010heatgreensfunction}:
\begin{equation}
    \begin{split}
        T_i(x,t) &= \theta_{\text{IC}}(x,0) + \theta_{\text{HG}}(x,t) + \theta_{\text{BCs}}(x,t) \\
        &= \sum_{j=1}^n  \int_{x_{j-1}}^{x_j} G_{i,j} (x,t|x',\tau)|_{\tau=0} T_{\text{IC}}(x')dx' \\
        &+ \sum_{j=1}^n \int_{\tau=0}^t \int_{x_{j-1}}^{x_j} G_{i,j}(x,t|x',\tau) \frac{\alpha_j}{k_j} \dot{q}_g(x',\tau)dx' d\tau \\
        &+ \left[\int_{\tau=0}^t G_{i,j}(x,t|x',\tau)|_{x'=0} \frac{\alpha_1}{k_1} f_1(x',\tau)  d\tau + \int_{\tau=0}^t G_{i,j}(x,t|x',\tau)|_{x'=x_n} \frac{\alpha_n}{k_n} f_n(x',\tau)  d\tau \right],
    \end{split}
    \label{eq:1D_GF_solution}
\end{equation}
where $\theta_{\text{IC}}$, $\theta_{\text{HG}}$, and $\theta_{\text{BCs}}$ represent the respective temperature contributions. $T_{\text{IC}}(x')$ denotes initial temperature at position $x'$, and $f_i(x',\tau)$ accounts for the time-dependent non-homogeneous boundary terms.
Physically, the 1D GF represents~\citep{carslaw1959conduction,davidw.hahnHeatConduction2012}: \textit{the temperature at location x, at time t, due to a unit instantaneous surface source at location x', at time $\tau$, as the medium being initialized at zero temperature and subjected to zero surface temperature.} 

Once the GF is established, the temperature distribution $T_i(x,t)$ can be calculated directly by Eq.~\eqref{eq:1D_GF_solution}. In this study, the GFs are derived based on the previously developed solution using the combined SOV-OE method, allowing for convenient implementation.

To construct the GFs, a simplified case is first considered: multi-layer composite with homogeneous cooling boundary condition ($T_{\text{water}} = 0$ K), and no internal heat generation ($\dot{q}_g(x',\tau) = 0$), where only the initial condition ($T_i(x,0)=T_{0}$) is present.
The solution for this case is easily obtained by the combined SOV-OE method by substituting Eq.~\eqref{eq:fourier coefficient} into Eq.~\eqref{eq:T_sp}, given as: 
\begin{equation}
T_i(x,t) =
\sum_{m=1}^\infty e^{-\lambda_m^2 t} \mathcal{X}_{i,m}(x) \frac{1}{N_m} \sum^n_{j=1}  \int^{x_j}_{x_{j-1}} \frac{k_j}{\alpha_j} T_{\text{IC}}(x')\mathcal{X}_{j,m}(x') dx',    
\label{eq. solution of the simplified case}
\end{equation}
where the $D_i$ and $E_i$ are zeros in this case. And Eq.~\eqref{eq:1D_GF_solution} can be simplified as:
\begin{equation}
\begin{split}
&T_i(x,t) = \theta_{\text{IC}}(x,0)\\
=& \sum^n_{j=1}  \int^{x_j}_{x_{j-1}} G_{i,j}(x,t|x',\tau)|_{\tau = 0} T_{\text{IC}}(x') dx'\\
=&\sum^n_{j=1}  \int^{x_j}_{x_{j-1}} \left[\sum_{m=1}^\infty  \frac{1}{N_m} \frac{k_j}{\alpha_j} e^{-\lambda_m^2 t} \mathcal{X}_{i,m}(x) \mathcal{X}_{j,m}(x')\right] T_{\text{IC}}(x') dx'.
\end{split}
\end{equation}
Compare Eq.~\eqref{eq:1D_GF_solution} with Eq.~\eqref{eq. solution of the simplified case}, the GFs evaluated at $\tau=0$ are then obtained, and the general form is found by replacing $t$ with $t-\tau$, as explained in~\citep{ozisik1993heat}, given as:
\begin{equation}
\begin{split}
 &G_{i,j}(x,t|x',\tau) \equiv G_{i,j}(x,t|x',\tau)|_{\tau = 0}, \ \text{for} \  t \rightarrow (t-\tau)\\
 & = \sum_{m=1}^\infty  \frac{1}{N_m} \frac{k_j}{\alpha_j} e^{-\lambda_m^2 (t-\tau)} \mathcal{X}_{i,m}(x) \mathcal{X}_{j,m}(x').
\end{split}
\label{eq:general GF}
\end{equation}
The temperature distribution is then obtained by substituting Eq.~\eqref{eq:general GF} into Eq.~\eqref{eq:1D_GF_solution}, incorporating all boundary and initial conditions and time-varying heat generation.

It is important to note that this solution is valid within the open interval $0<x<x_n$, due to the non-uniform convergence of series at the boundaries $x=0, x_n$, in Eq.~\eqref{eq:1D_GF_solution}. This issue arises when using basis functions derived from the OE technique, where the non-homogeneous boundary conditions are incorporated into the eigenvalue problem, as also presented in~\citep{davidw.hahnHeatConduction2012}.
This limitation can be addressed by eliminating non-homogeneities from the boundary conditions through the superposition method~\citep{ozisik1993heat}, though this procedure is not detailed here.

\subsection{Sinusoidal and square wave heat sources}

\begin{figure*}[h!]
\centering
\begin{subfigure}{0.48\textwidth}
    \includegraphics[width=\linewidth]{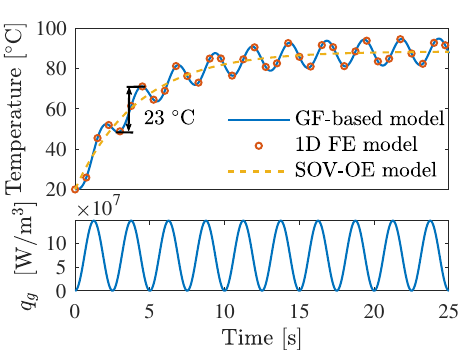}
    \caption{Sinusoidal heat generation}
    \label{fig:Sinusoidal}
\end{subfigure}
\hfill
\begin{subfigure}{0.48\textwidth}
    \includegraphics[width=\linewidth]{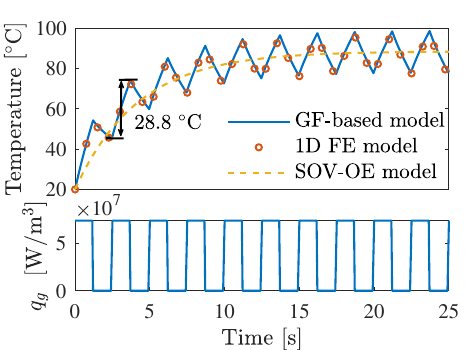}
    \caption{Square wave heat generation}
    \label{fig:Step wave}
\end{subfigure}
\caption{Comparison of the temporal temperature profiles over 25 s at the enamel-epoxy interface ($x=x_2$). Results are obtained from: (1) the GF-based model with dynamic heat sources; (2) the 1D FE model with identical dynamic heat sources; and (3) the combined SOV-OE method with an equivalent constant heat source calculated from the RMS value of the current.}
\label{fig:transient sources_sin and step}
\end{figure*}

The GF-based model enables fast and continuous analytical solutions for the temperature distribution in the multi-layer composites subjected to time-varying heat sources. To validate this approach, two types of current waveforms were applied: a 0.4 Hz sinusoidal current and a square wave with the same frequency and root mean square (RMS) amplitude, generating corresponding ohmic losses in the coil layer. The temperature at the enamel-epoxy interface was monitored for comparison. Temperature profiles obtained using the GF-based method were compared with those from a 1D FE model subjected to identical time-varying heat generation. Additionally, a reference profile was generated using the combined SOV-OE model, with a constant $\dot{q}_g$ calculated from the RMS value of the dynamic current. The results, presented in Fig~\ref{fig:transient sources_sin and step}, show strong agreement between the GF-based and 1D FE models. Simulation with the GF-based model took approximately 1 second, in contrast to 30 seconds needed by the 1D FE model under the same conditions. 

Although the average temperature rise under time-varying and constant heat sources is similar, temperature fluctuations under dynamic conditions result in distinct thermal effects. These fluctuations can accelerate thermal aging and fatigue, ultimately reducing the lifetime of the materials. Notably, the temperature variations are more pronounced under square wave current compared to the sinusoidal wave, due to the abrupt changes in heat generation. This has meaningful implications for reliability and lifetime assessment of components subjected to dynamic thermal loads, such as the PMLSM examined in this research.
Moreover, temperature fluctuations diminish at higher frequencies, with the temperature profile gradually converging to that of the constant $\dot{q}_g$ case. Determining an optimal operating frequency based on the temperature response can contribute to improve the composite design and prolong the service life of multi-layer composites. 

\subsection{Heat source from motion profile}

\begin{figure*}[h!]
    \centering
    \includegraphics[width=0.75\linewidth]{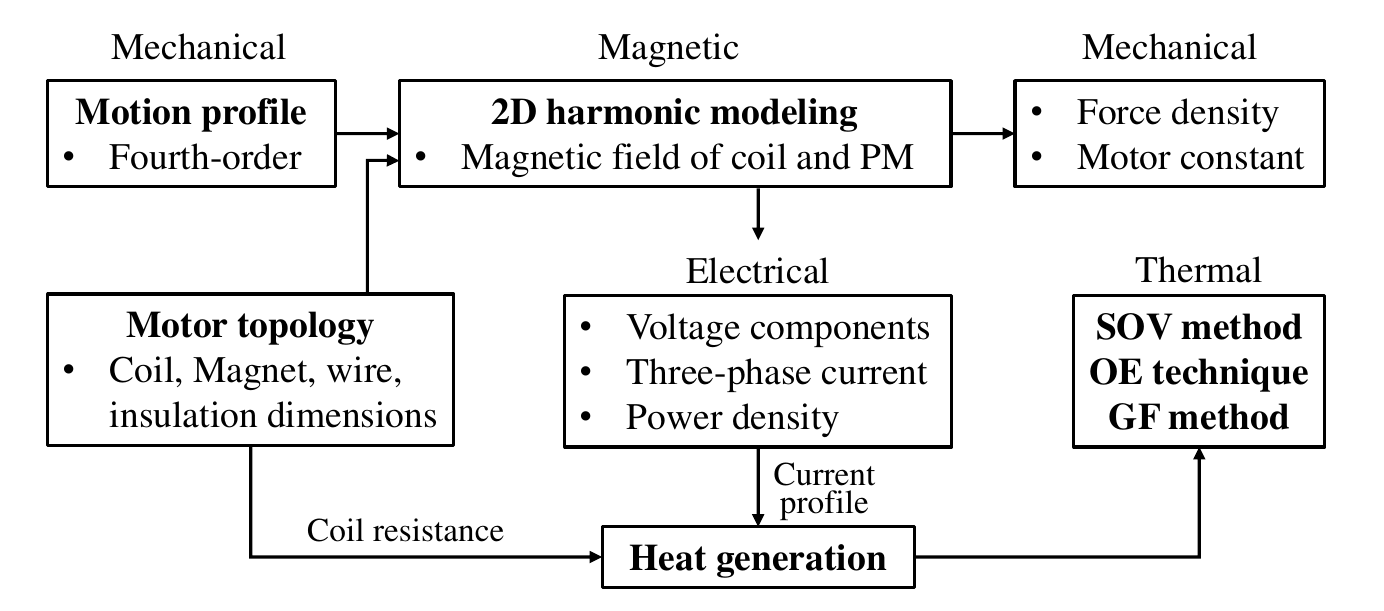}
    \caption{Schematic of the coupled multi-physics PMLSM model.}
    \label{fig:multiphysics model}
\end{figure*}

In certain applications, multi-layer composites are subjected to highly dynamic heat generation. For example, in the lithographical industry, wafer stages require stable, precise, and rapid motion, which results in dynamically varying current excitation in the driving motors. To illustrate the practical application of the GF-based approach, a multi-physics PMLSM model~\citep{roversMultiphysicalModelingHighPrecision2013}, as depicted in Fig.~\ref{fig:multiphysics model}, is coupled with the GF-based forward thermal model. The thermal behavior is evaluated under a designed motion profile, representing realistic heat generation dynamics.

\begin{figure*}[h!]
\centering
\begin{subfigure}{0.48\textwidth}
    \includegraphics[width=\linewidth]{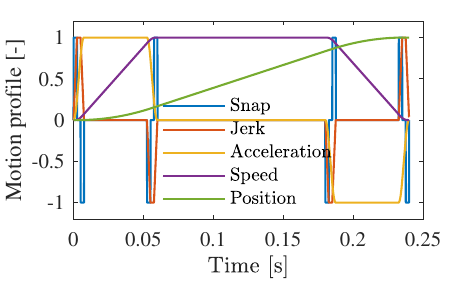}
    \caption{Normalized fourth-order motion profile.}
    \label{fig:motion profile}
\end{subfigure}
\hfill
\begin{subfigure}{0.48\textwidth}
    \includegraphics[width=\linewidth]{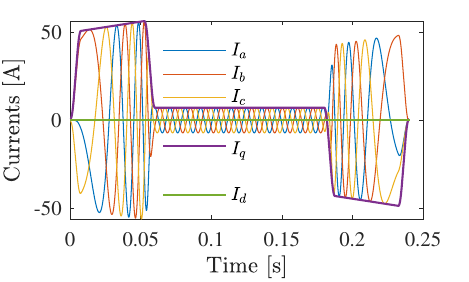}
    \caption{Required current profile.}
    \label{fig:current profile}
\end{subfigure}
    \caption{Motion profile and corresponding current profile of the PMLSM over one period.}
    \label{fig:current and motion profile}
\end{figure*}

\begin{figure*}
    \centering
    \includegraphics[width=1\linewidth]{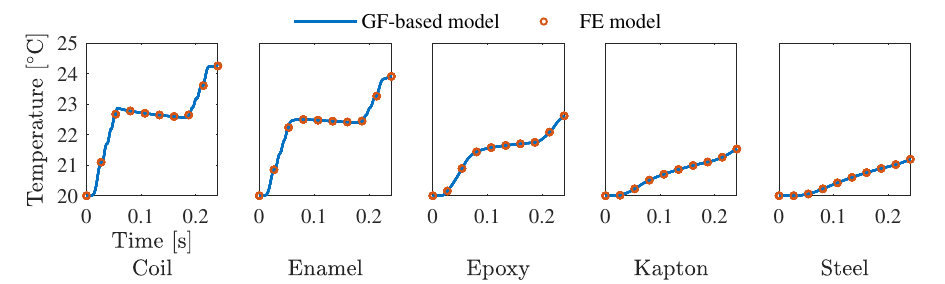}
    \caption{Comparison of temperature profiles at the centers of all five layers (Coil, enamel, epoxy, Kapton, steel) during one motion period, obtained from the GF-based model and the 1D FE model.}
    \label{fig:T under motion profile}
\end{figure*}

This multi-physics model integrates the mechanical performance with electrical inputs through magnetic field calculations. 
A representative fourth-order trapezoidal motion profile~\citep{settelsFastSwitchingHigh2019}, consisting of acceleration, constant speed, and deceleration phases, is defined as one motion period in Fig.~\ref{fig:motion profile}.
Under a specific motor topology, the current required to follow the designed motion profile is computed by a 2D Harmonic modeling approach~\citep{gysen2010general,jansen2014overview}. From the resulting currents profile, as shown in Fig.~\ref{fig:current profile}, the internal heat generation $\dot{q}_g$ in the coil layer is calculated and incorporated into the GF-based forward thermal model to determine the temperature distribution. 
To explore thermal dynamics across different layers, temperature profiles at the center of each of the five layers are plotted over one period of the current profile, as shown in Fig.~\ref{fig:T under motion profile}.
The results show that the coil layer responds more rapidly to variations in heat generation due to its larger thermal diffusivity. As heat propagates through the layers to the cooling water, high frequency components are dampened. This attenuation is due to the lower thermal diffusivity and large time constants of the polymer layers. Additionally, the layers between the coil and the cooling water act as `low-pass thermal filters', smoothing out fluctuations from the heat source. 

\begin{figure}[h!]
    \centering
        \includegraphics[width=0.48\textwidth]{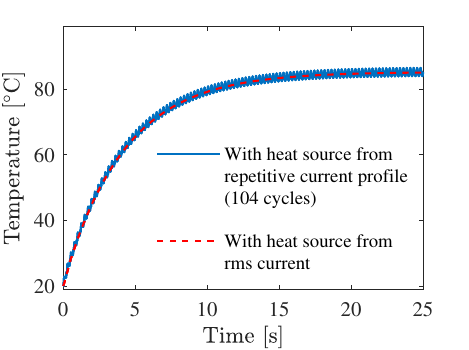}
    \caption{Temperature rise at enamel-epoxy interface over 25 s under a dynamic heat source from 104 repeated motion cycles, compared with an equivalent constant heat source.}
    \label{fig:T motion profile for 25 s}
\end{figure}

To assess the thermal behavior in a longer time period, the motion profile is repeated over 104 cycles, and the resulting temperature response at the enamel-epoxy interface is shown in Fig.~\ref{fig:T motion profile for 25 s}, along with the temperature profile obtained from the combined SOV-OE method with a constant $\dot{q}_g$ calculated from the RMS value of the current profile. 

The results indicate that this GF-based approach remains robust over extended durations, while detailed temperature variations are captured.
Temperature fluctuations are influenced not only by the variation in the heat generation, but also by material properties and layer geometry. This is critical for evaluating whether a composite design is suitable for specific thermal loads. 
This example highlights the flexibility of the GF-based modeling approach and its potential as a tool for analyzing heat propagation in multi-layer composite structures. 

\subsection{Temperature-dependent resistivity}
A further extension of the GF-based model addresses the temperature-dependent electrical resistivity of copper. Since the ohmic losses are the sole source of heat generation in this example, both the time-varying current and temperature-dependent resistivity affect the temperature distribution. 
Directly solving the heat diffusion equation with $\dot{q}_g(T, t)$ introduces non-linearity, significantly increasing computational cost and complicating the use of the GF-based approach.

To address this, a fixed-point iteration approach is employed: where the coil resistance is updated using the GF-based model. Leveraging the merit of this GF-based approach, the temperature profile is computed over the entire time span instead of by time steps. The iterative process involves:

1. Solving for the initial temperature profile using a prescribed $\dot{q}_{g,1}(t)$.

2. Updating the coil resistance based on this temperature profile, yielding a new $\dot{q}_{g,2}(t)$.

3. Recomputing the temperature profile and comparing it to the previous iteration.

4. Repeating the process until the error falls below a defined tolerance $\varepsilon$.

\begin{figure}[h!]
    \centering
    \includegraphics[width=0.48\textwidth]{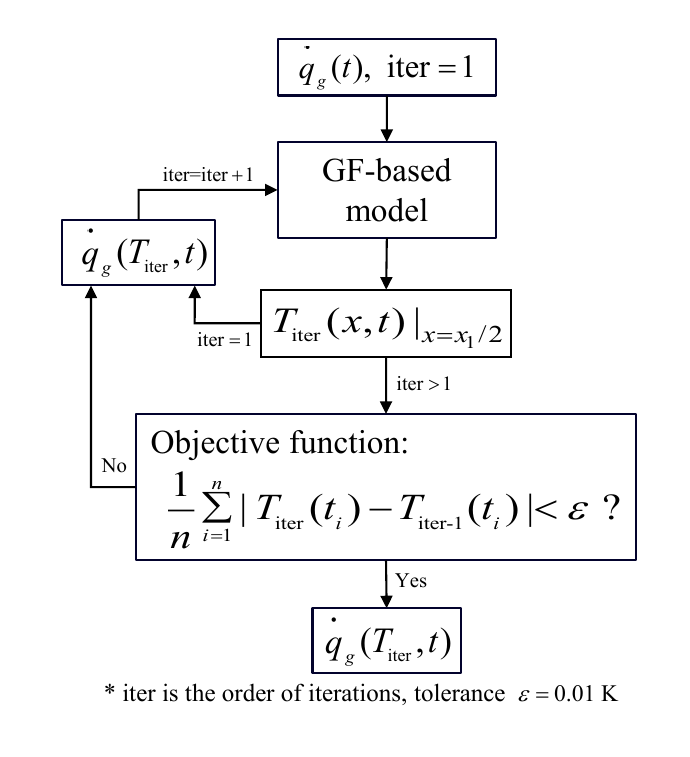}
    \caption{Schematic of the fixed-point iteration process for incorporating temperature-dependent copper resistivity.}
    \label{fig:flowchat_TdependentR}
\end{figure}

As shown in Fig.~\ref{fig:flowchat_TdependentR}, and executed with the GF-based model.

\begin{figure*}[h!]
\centering
\begin{subfigure}{0.48\textwidth}
    \includegraphics[width=\linewidth]{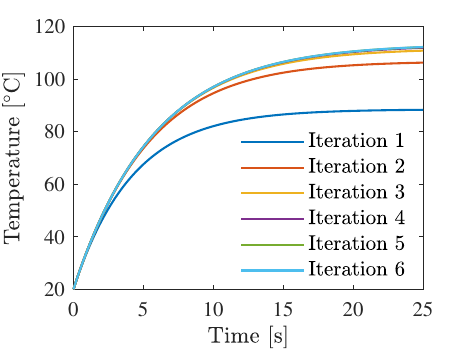}
    \caption{Temperature profile $T_k(t)$ at the center of coil layer of iteration $k=1$ to $6$, with tolerance of $\varepsilon$= 0.01 $\mathrm{K}$.}
    \label{fig:T_plot varying resistivity}
\end{subfigure}
\hfill
\begin{subfigure}{0.48\textwidth}
    \includegraphics[width=\linewidth]{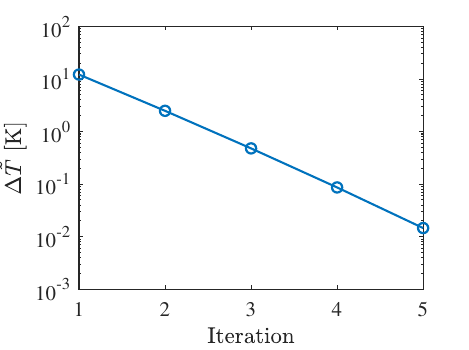}
    \caption{Convergence of the average temperature variation during iterative updates of coil resistance and heat generation.}
    \label{fig:varying resistivity}
\end{subfigure}
    \caption{Temperature profiles with temperature-dependent copper resistivity.}
    \label{fig:current and motion profile}
\end{figure*}

Indicated in Fig.~\ref{fig:T_plot varying resistivity}, a significant difference of approximately 15 $\mathrm{K}$ was observed between the steady-state temperatures from the first and the final iterations. 

The objective function, representing the difference between iterations, converges exponentially and reaches the tolerance threshold (0.01 $\mathrm{K}$) after just six iterations, as shown in Fig.~\ref{fig:varying resistivity}. This iterative process adds minimal computational costs since the eigenvalues $\lambda$ and other coefficients remain unchanged throughout.

To explore thermal dynamics across different materials, temperature profiles at the center of each of the five layers are plotted over one period of the current profile, as shown in Fig.~\ref{fig:T under motion profile}.

The results in this section, not only highlight the effectiveness of the GF-based approach in handling dynamic thermal scenarios but also provide a preliminary forward thermal model to incorporate into an inverse problem framework for parameter estimation, which will be further explored in Part II of this research.





\section{Conclusions}
This paper, as the first part of a two-part study, presents fast and accurate analytical approaches for modeling heat conduction in multi-layer composites subjected to both constant and time-varying heat generation. These methods yield exact closed-form solutions for transient temperature distribution, providing a reliable and computationally efficient alternative to numerical methods.
A detailed implementation of a combined approach using the SOV method and OE technique has been developed. This approach is general and applicable to composites with an arbitrary number of layers under constant internal heat generation. 
Validation against FE models confirms the high accuracy of the proposed models, with significantly reduced computation time.

To address scenarios involving dynamic heat sources and temperature-dependent copper resistivity, a Green’s function-based method is introduced, building on the foundation of the SOV-OE solution. This approach effectively captures transient thermal behavior without the need for iterative coefficient updates at each time step.
A practical application involving a multi-physics PMLSM model demonstrates how dynamic heat generation derived from motion profiles can be accurately modeled using this method.
Furthermore, an iterative scheme is proposed to incorporate temperature-dependent electrical resistivity, enabling precise thermal predictions under realistic conditions.

Overall, these forward models enable efficient and accurate evaluation of temperature distributions in multi-layer composites under various working conditions, and are particularly suited for integration into portable thermal evaluation systems where fast response is critical. Importantly, they also serve as the foundation for solving real-time inverse heat conduction problems, providing a robust tool for material parameter estimation. The application of these models within an inverse framework is explored in detail in Part II of this research.

\printcredits

\bibliographystyle{elsarticle-num_nourl}

\bibliography{references}
\appendix
\clearpage
\clearpage
\renewcommand{\theequation}{A.\arabic{equation}}
\setcounter{equation}{0} 

\nomenclature{$L$}{Linear operator}
\nomenclature{$L^*$}{Modified linear operator, with a weight of $k_i$}

\section{Natural orthogonality}\label{Appendix_OET}
The spatial component $\mathcal{X}_i(x)$ forms a \textit{Helmholtz equations}, written as:
\begin{equation}
    L[\mathcal{X}_{i}] = \mathcal{X}_i^{''}(x) = -\frac{\lambda^2}{\alpha_i} \mathcal{X}_i(x),
    \label{eq.Helmholtz}
\end{equation}
where $L$ is the linear operator. Subjected to the boundary conditions, this problem constitutes a regular Sturm-Liouville boundary value problem~\citep{braunDifferentialEquationsTheir1993}, where $L$ must be self-adjoint. Let $\mathcal{X}_{i,m}$ and $\mathcal{X}_{i,p}$ be eigenfunctions corresponding to distinct eigenvalues $\lambda_m$ and $\lambda_p$ ($\lambda_m \neq \lambda_p$). The self-adjointness condition requires: 
\begin{equation}
    (L \mathcal{X}_{i,m}, \mathcal{X}_{i,p}) = (\mathcal{X}_{i,m}, L \mathcal{X}_{i,p}),
\label{eq:self-adjointness}
\end{equation}
with the inner product defined as:
\begin{equation}
    (u,v) = \int_a^b u(x)v(x)dx.
\end{equation}
Expanding Eq.~\eqref{eq:self-adjointness}, we obtain:
\begin{equation}
\begin{split}
       &  \int_0^{x_n} (\mathcal{X}_{i,m}^{''} \mathcal{X}_{i,p} - \mathcal{X}_{i,p}^{''} \mathcal{X}_{i,m}) dx = [\mathcal{X}_{i,m}^{'} \mathcal{X}_{i,p} - \mathcal{X}_{i,p}^{'} \mathcal{X}_{i,m}]|_0^{x_n} \\
        =& [\mathcal{X}_{1,m}^{'} \mathcal{X}_{1,p} - \mathcal{X}_{1,p}^{'} \mathcal{X}_{1,m}]|_0^{x_1}+\sum_{i=2}^n [\mathcal{X}_{i,m}^{'} \mathcal{X}_{i,p} - \mathcal{X}_{i,p}^{'} \mathcal{X}_{i,m}]|_{x_{i-1}}^{x_i} \\
       =& \sum_{i=1}^n \int_0^{x_n} -\frac{(\lambda_{m}^2-\lambda_{p}^2)}{\alpha_i} \mathcal{X}_{i,m} \mathcal{X}_{i,p} dx = 0.
\end{split}
\label{eq:self-adjointness_expanded}
\end{equation}
Since all eigenfunctions satisfy the boundary conditions, the cooling and symmetrical boundary terms vanish, leaving only the continuous boundary terms to be considered. The equilibrium of Eq.~\eqref{eq:self-adjointness_expanded} is then simplified as:
\begin{equation}
\begin{split}
        \mathcal{X}_{i,m}^{'} (x_i) \mathcal{X}_{i,p}(x_i) - \mathcal{X}_{i+1,m}^{'} (x_i) \mathcal{X}_{i+1,p}(x_i) &= 0, \\
        \mathcal{X}_{i,p}^{'}(x_i) \mathcal{X}_{i,m}(x_i)  - \mathcal{X}_{i+1,p}^{'}(x_i) \mathcal{X}_{i+1,m}(x_i) & = 0. \\
        (i=1,...,n-1)&
        \label{eq.ortho_2}
\end{split}
\end{equation}
Considering the continuous boundary conditions:
\begin{equation}
    \begin{split}
        \mathcal{X}_{i,m}(x_i) &=  \mathcal{X}_{i+1,m}(x_{i}), \\
        k_i \mathcal{X}_{i,m}^{'}(x_i) &=   k_{i+1} \mathcal{X}_{i+1,m}^{'}(x_{i}). \\
        (i&=1,...,n-1)
    \end{split}
\end{equation}
It becomes evident that a weighting factor $k_i$ is required to preserve self-adjointness, yielding a modified Helmholtz equation:
\begin{equation}
    L^*[\mathcal{X}_{i}] = k_i \mathcal{X}_i^{''}(x) = -\lambda^2 \frac{k_i}{\alpha_i} \mathcal{X}_i(x),
    \label{eq.Helmholtz_new}
\end{equation}
where $L^*$ is the modified linear operator. A fundamental theorem in the regular Sturm-Liouville problem states that~\citep{braunDifferentialEquationsTheir1993}: \textit{Eigenfunctions belonging to different eigenvalues are orthogonal under the inner product}:
\begin{equation}
    \langle u,v\rangle = \int_a^b r(x) u(x)v(x)dx,
\end{equation}
where, in this case, $r(x) = k_i/\alpha_i$. 

Another theorem in the regular Sturm-Liouville problem states~\citep{braunDifferentialEquationsTheir1993}:
\textit{For any continuously differentiable function $F_i(x)$ on the interval $[0,x_n]$. $F_i(x)$ can be expanded in a convergent series of the eigenfunctions of $L$}; i.e.,
\begin{equation}
    F_i(x) = T_0 - D_i x - E_i = \sum_{m=1}^\infty C_m  \mathcal{X}_{i,m}(x).
    \label{eq:SL_theorem2}
\end{equation}

The Fourier coefficients $C_m$ can be derived by exploiting the self-adjointness of the eigenfunctions. Considering the inner product of $L^* \mathcal{X}_{i,m}$ with $\mathcal{X}_{i,p}$:
\begin{equation}
\begin{split}
    (L^* \mathcal{X}_{i,m}, \mathcal{X}_{i,p}) &= (-\lambda_m^2 \frac{k_i}{\alpha_i} \mathcal{X}_{i,m}, \mathcal{X}_{i,p}) \\
    &=-\lambda_m^2 \int_0^{x_n} \frac{k_i}{\alpha_i} \mathcal{X}_{i,m}(x) \mathcal{X}_{i,p}(x) dx.
\end{split}
\end{equation}
Since $L^*$ is self-adjoint:
\begin{equation}
\begin{split}
    (L^* \mathcal{X}_{i,m}, \mathcal{X}_{i,p}) &=(\mathcal{X}_{i,m}, L^*  \mathcal{X}_{i,p}) \\
    &=(\mathcal{X}_{i,m}, -\lambda_m^2 \frac{k_i}{\alpha_i} \mathcal{X}_{i,p}) \\
    &=-\lambda_p^2  \int_0^{x_n} \frac{k_i}{\alpha_i} \mathcal{X}_{i,m}(x) \mathcal{X}_{i,p}(x) dx.
\end{split}
\end{equation}
For $\lambda_m \neq \lambda_p$, it follows that:
\begin{equation}
   \int_0^{x_n} \frac{k_i}{\alpha_i} \mathcal{X}_{i,m}(x) \mathcal{X}_{i,p}(x) dx = \langle\mathcal{X}_{i,m},\mathcal{X}_{i,p}\rangle =0.
\end{equation}
This property is referred to as `natural' orthogonality~\citep{demonteTransientHeatConduction2000,demonteAnalyticApproachUnsteady2002, demonte2003unsteady}. 

Taking the inner product of $F_i(x)$ with any eigenfunction $\mathcal{X}_{i,m}(x)$:
\begin{equation}
\begin{split}
\langle F_i, \mathcal{X}_{i,m}\rangle &=  \sum_{p=1}^\infty C_p \int_0^{x_n} \mathcal{X}_{i,m}(x) \mathcal{X}_{i,p}(x) dx  \\
&=\sum_{p=1}^\infty C_p \langle \mathcal{X}_{i,m},\mathcal{X}_{i,p}\rangle \\
&= C_m \langle\mathcal{X}_{i,m},\mathcal{X}_{i,m}\rangle.
\end{split}
\end{equation}
The only remaining unknown $C_m$ is thus given by:
\begin{equation}
    \begin{split}
        C_m &= \frac{\langle F_i, \mathcal{X}_{i,m}\rangle}{\langle \mathcal{X}_{i,m},\mathcal{X}_{i,m}\rangle} \\
        &= \frac{1}{N_m}\sum^n_{i=1}  \int^{x_i}_{x_{i-1}} \frac{k_i}{\alpha_i} F_i(x)\mathcal{X}_{i,m}(x) dx,
    \end{split}
\end{equation}
where $N_m = \langle \mathcal{X}_{i,m},\mathcal{X}_{i,m} \rangle$ is the norm.



\end{document}